\documentclass[conference]{IEEEtran}
\IEEEoverridecommandlockouts
\usepackage{cite}
\usepackage{amsmath,amssymb,amsfonts}
\usepackage{algorithmic}
\usepackage{graphicx}
\usepackage{textcomp}
\usepackage{xcolor}

\usepackage{kotex}
\usepackage{hyperref}

\usepackage{multirow}
\usepackage{booktabs}

\usepackage[numbers,sort&compress]{natbib}

\hypersetup{
}

\def\BibTeX{{\rm B\kern-.05em{\sc i\kern-.025em b}\kern-.08em
    T\kern-.1667em\lower.7ex\hbox{E}\kern-.125emX}}

\makeatletter
\newcommand{\linebreakand}{%
  \end{@IEEEauthorhalign}
  \hfill\mbox{}\par
  \mbox{}\hfill\begin{@IEEEauthorhalign}
}
\makeatother

\begin{document}


\title{QR-VC: Leveraging Quantization Residuals for Linear Disentanglement in Zero-Shot Voice Conversion}




%


\author{\IEEEauthorblockN{Youngjun Sim\IEEEauthorrefmark{1}}
\IEEEauthorblockA{\textit{Graduate School of Artificial}\\ \textit{Intelligence} \\
\textit{POSTECH}\\
Pohang, South Korea\\
youngjunsim@postech.ac.kr}
\and
\IEEEauthorblockN{Jinsung Yoon\IEEEauthorrefmark{1}}
\IEEEauthorblockA{\textit{Graduate School of Artificial}\\ \textit{Intelligence} \\
\IEEEcompsocitemizethanks{\IEEEcompsocthanksitem\IEEEauthorrefmark{1}Equal contribution}
\textit{POSTECH}\\
Pohang, South Korea\\
truestar2001@postech.ac.kr}
\and
\IEEEauthorblockN{Wooyeol Jeong}
\IEEEauthorblockA{\textit{Graduate School of Artificial}\\ \textit{Intelligence} \\
\textit{POSTECH}\\
Pohang, South Korea\\
jungwy0106@postech.ac.kr}
\and
\IEEEauthorblockN{Young-Joo Suh\IEEEauthorrefmark{2}}
\IEEEcompsocitemizethanks{\IEEEcompsocthanksitem\IEEEauthorrefmark{2}Corresponding author}
\IEEEauthorblockA{\textit{Graduate School of Artificial}\\ \textit{Intelligence} \\
\textit{POSTECH}\\
Pohang, South Korea\\
yjsuh@postech.ac.kr}
}

\maketitle

\begin{abstract}
Zero-shot voice conversion is a technique that alters the speaker identity of an input speech to match a target speaker using only a single reference utterance, without requiring additional training.
Recent approaches extensively utilize self-supervised learning features with K-means quantization to extract high-quality content representations while removing speaker identity. However, this quantization process also eliminates fine-grained phonetic and prosodic variations, degrading intelligibility and prosody preservation.
While prior works have primarily focused on quantized representations, quantization residuals remain underutilized and deserve further exploration. 
In this paper, we introduce a novel approach that fully utilizes quantization residuals by leveraging temporal properties of speech components. 
This facilitates the disentanglement of speaker identity and the recovery of phonetic and prosodic details lost during quantization.
By applying only K-means quantization and linear projections, our method achieves simple yet effective disentanglement, without requiring complex architectures or explicit supervision. 
This allows for high-fidelity voice conversion trained solely with reconstruction losses.
Experiments show that the proposed model outperforms existing methods across both subjective and objective metrics.
It achieves superior intelligibility and speaker similarity, along with improved prosody preservation, highlighting the impact of our Linear Disentangler module.
\end{abstract}


\section{Introduction}
\label{sec:intro}
Zero-shot voice conversion (VC) aims to transform the speaker identity of a source speech into that of an arbitrary target using only a single utterance, without additional training. Generating high-fidelity converted speech requires a distinct disentanglement of content-related and speaker-specific components while preserving intelligibility, prosody, and speaker similarity.

Self-supervised learning (SSL) models~\cite{hubert, wav2vec2, wavlm} have gained significant attention in VC research as they are trained to encode latent speech representations.
The resulting SSL features have been shown to linearly predict various speech attributes~\cite{wavlm, ssl_layer, ssl_prosody}, suggesting that key components of speech are linearly separable.
This separability underpins the design of our linear disentanglement module.

Over the past few years, zero-shot VC performance has been greatly enhanced by integrating SSL features with techniques such as diffusion models~\cite{dddm-vc}, conformer-based architectures~\cite{sef-vc}, normalizing flows~\cite{free-vc}, explicit supervision like prosodic labels~\cite{ssr-vc, sef-vc}, and the use of external speaker encoders~\cite{ssr-vc, free-vc}, within a disentanglement module.
Nevertheless, these approaches typically rely on complex deep architectures, resulting in increased model size and computational overhead.

One key property of SSL features is that nearby features share phonetic information~\cite{knn-vc, ssl_phoneme}. Leveraging this property, K-means quantization (KQ) has been widely adopted as a strong content encoder in various VC models~\cite{ssr-vc, sef-vc, vec-tok}. This process effectively removes speaker information while capturing essential content representations.

However, quantization eliminates fine-grained phonetic and prosodic variations, referred to as speaking variations, leading to intelligibility degradation and poor prosody preservation. Existing models struggle to reconstruct these lost details, suggesting that quantization residuals are worth exploring further. 

We propose QR-VC, a novel zero-shot VC model that leverages quantization residuals and the temporal properties of speech components to construct a simple yet effective architecture. Our model disentangles phonetic content, speaker identity, and speaking variations using only K-means quantization and linear projection layers. This approach enables high-fidelity VC relying solely on reconstruction losses, without requiring explicit supervision or complex modules.
In particular, the proposed Linear Disentangler module, composed of time-invariant and time-variant linear bottlenecks, addresses the limitations of quantization by restoring fine-grained speaking variations, even with a small codebook.
Through extensive experiments, we demonstrate that QR-VC outperforms existing methods in terms of intelligibility, speaker similarity, prosody preservation, and naturalness. 
Our results highlight the importance of quantization residuals and validate the effectiveness of a linear disentanglement framework in zero-shot VC.
Audio samples are available at \href{https://simyoungjun.github.io/QR-VC-DEMO/}{\texttt{https://simyoungjun.github.io/QR-VC-DEMO/}}

\setlength{\tabcolsep}{6pt}
\begin{table*}[!ht]
\centering
\normalsize
\caption{Subjective and objective evaluation results for MOS, SMOS, WER, CER, EER, SECS, E-PCC, and F0-PCC.}

\begin{tabular}{lcccccccc} 
\toprule
\multicolumn{1}{c}{Model} & MOS$\uparrow$ & SMOS$\uparrow$ & WER(\%)$\downarrow$ & CER(\%)$\downarrow$ & EER(\%)$\downarrow$ & SECS$\uparrow$ & E-PCC$\uparrow$ & F0-PCC$\uparrow$ \\
\midrule
SSR-VC \cite{ssr-vc} & 2.97$\pm$0.17 & 3.04$\pm$0.16 & 11.80 & 4.64 & 18.6 & 0.725 & 0.701 & 0.692 \\
SEF-VC \cite{sef-vc} & 3.98$\pm$0.14 & 3.41$\pm$0.15 & 7.57 & 2.47 & 16.6 & 0.707 & 0.708 & 0.657 \\
Phoneme Hallucinator \cite{phoneme_hallu} & 3.81$\pm$0.18 & 3.76$\pm$0.13 & 10.78 & 4.76 & 12.8 & 0.743 & 0.793 & 0.716 \\
DDDM-VC \cite{dddm-vc} & 3.69$\pm$0.20 & 4.03$\pm$0.18 & 12.05 & 5.19 & 8.0 & 0.791 & 0.685 & 0.685 \\
kNN-VC \cite{knn-vc} & 2.78$\pm$0.15 & 3.68$\pm$0.19 & 37.58 & 20.81 & \textbf{4.0} & 0.798 & 0.622 & 0.628 \\
\midrule
\textbf{QR-VC (Ours)} & \textbf{4.14$\pm$0.12} & \textbf{4.27$\pm$0.13} & \textbf{5.57} & \textbf{2.07} & 5.0 & \textbf{0.815} & \textbf{0.885} & \textbf{0.719} \\
\bottomrule
\end{tabular}
\label{table_1}
\end{table*}

\section{Method}
\begin{figure}[!tb]
\begin{minipage}[b]{1.0\linewidth}
  \centering
  \centerline{\includegraphics[width=9cm]{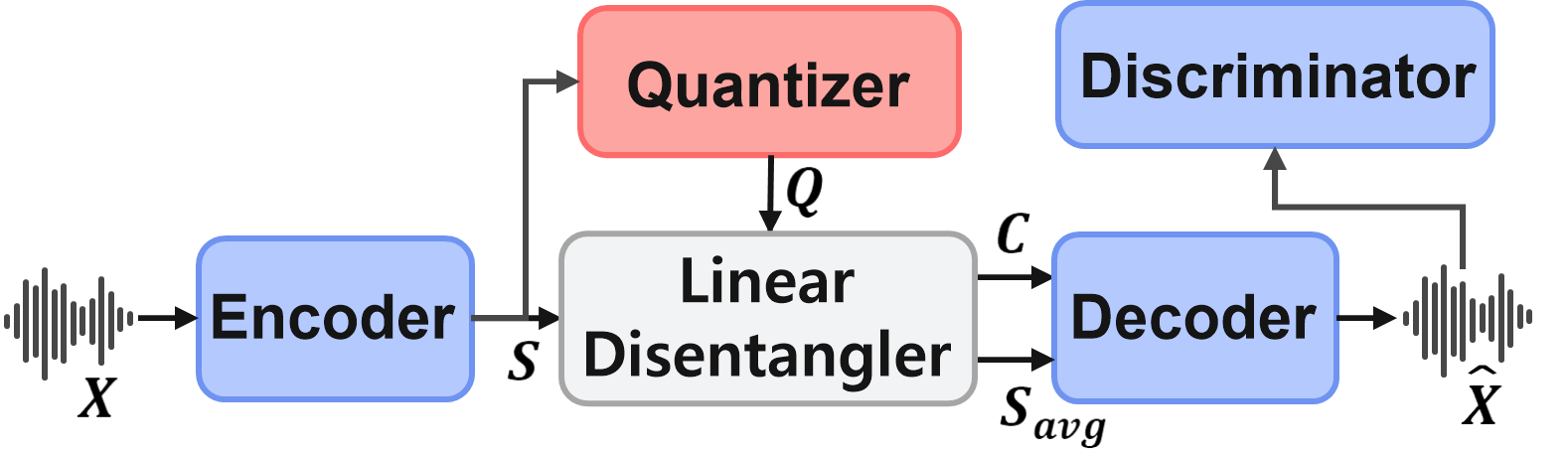}}
  \footnotesize\centerline{(a) QR-VC}\medskip
\end{minipage}
\begin{minipage}[b]{1.0\linewidth}
  \centering
  \centerline{\includegraphics[width=8.6cm]{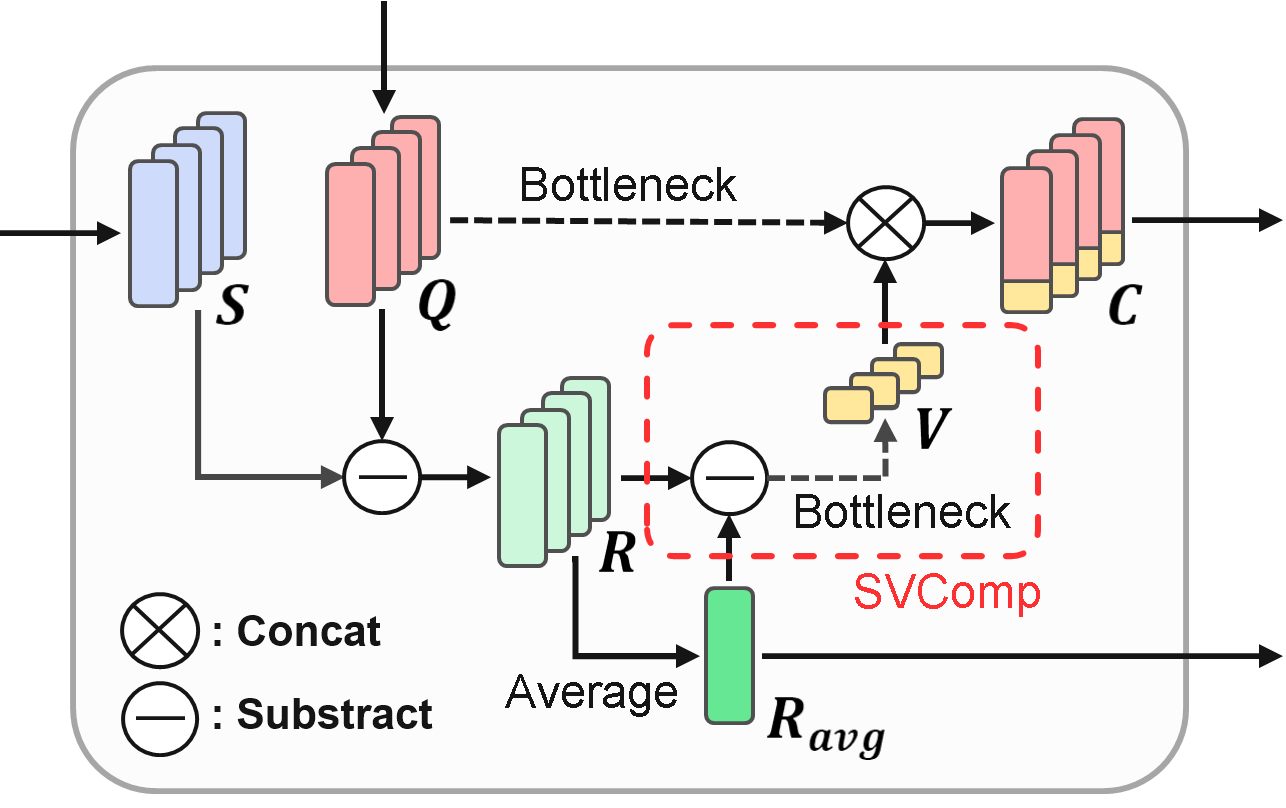}}
  \footnotesize\centerline{(b) Linear Disentangler}\medskip
\end{minipage}
\caption{Overview of QR-VC. (a) Overall architecture. (b) Detailed structure of the Linear Disentangler module.}
\label{model_train}
\end{figure}
\subsection{Overview}
QR-VC consists of a WavLM~\cite{wavlm}-based encoder, a Linear Disentangler, and a HiFi-GAN~\cite{hifi} decoder. The key improvement lies in leveraging quantization residuals to disentangle speaker identity and speaking variations through a linear structure. This approach preserves fine-grained prosodic and phonetic details while ensuring speaker similarity.

\subsection{Encoder and Quantizer}
The encoder extracts SSL features $S$ from the \(6^{th}\) layer of WavLM given a speech input \( X \), defined as:
\begin{equation}
    S = [s_1, s_2, ..., s_T] \in \mathbb{R}^{1024 \times T}.
\end{equation}
where \(T\) denotes the number of frames. Previous studies have demonstrated that these features, in particular, retain rich phonetic, prosodic, and speaker information~\cite{knn-vc, ssl_prosody}.

To obtain a discrete content representation $Q$, we apply K-means quantization using a codebook $E$:
\begin{equation}
    \mathbf{E} = \{e_1, e_2, ..., e_K\} \in \mathbb{R}^{K \times 1024}.
\end{equation}
where \(K\) is the number of codebook entries. Each frame \( s_t \) is mapped to the nearest entry:
\begin{equation}
    q_t = \arg\min_{e_k \in E} \|s_t - e_k\|_2^2,\; Q = [q_1, q_2, ..., q_T].
\end{equation}
Quantization removes speaker identity while preserving essential linguistic content. However, fine-grained speaking variations, such as phonetic and prosodic details, are also lost.
\subsection{Linear Disentangler}
To recover these lost details, we introduce a disentanglement mechanism that decomposes the quantization residual $R$:
\begin{equation}
    r_t = s_t - q_t, \; R \in \mathbb{R}^{1024 \times T}.
\end{equation}
This quantization residual $R$ preserves fine-grained phonetic and prosodic details, along with speaker information, as shown in Figure~\ref{feature_analysis}. The disentangler further separates these components into time-invariant speaker identity and time-varying speaking variations. 

\subsubsection{Speaker Identity Extraction}
Since speaker identity is generally consistent across an utterance, we extract it by computing the temporal average of \(R\):
\begin{equation}
    R_{avg} = \frac{1}{T} \sum_{t=1}^{T} r_t.
\end{equation}
This removes time-varying fluctuations while preserving time-invariant speaker identity characteristics.

\subsubsection{Speaking Variation Compensation (SVComp)}
To isolate speaking variations \( V \), we subtract the speaker identity vector \( R_{\text{avg}} \) from each quantization residual frame $r_t$. 
The result is passed through a \(1 \times 1\) convolutional bottleneck layer, which serves as a linear projection. This operation captures important time-varying information:

\begin{equation}
    V = \text{Bottleneck}(R - R_{avg}),\; v_t \in \mathbb{R}^{M}.
\end{equation}
where \( M \) is the bottleneck dimension. The resulting representation \( V \) retains prosodic and phonetic details while suppressing speaker information.

\subsubsection{Final Content Representation}
\label{sec:final_content_representation}
The disentangled content representation \( C \) is obtained by concatenating the bottlenecked quantized representation $Q'$ with the extracted speaking variations \( V \). The two components have dimensions of \((1024 - M) \times T\) and \( M \times T \), respectively, forming a 1024-dimensional representation:
\begin{equation}
    C = Q' \mathbin{\otimes} V,\; c_t \in \mathbb{R}^{1024}.
\end{equation}
This process compensates for the loss of speaking variations in quantization, leading to improved content and prosody preservation.

\subsection{Decoder and Training Objective}
\label{sec:Training Objective}
The decoder $G$ reconstructs the speech waveform \(\hat{X}\) using HiFi-GAN~\cite{hifi} generator, taking the concatenated content \( C \) and speaker identity $R_{avg}$ as the input:
\begin{equation}
    \hat{X} = G(C + R_{avg}).
\end{equation}
The model is trained with a combination of adversarial losses $L_{adv}(G)$ and $L_{adv}(D)$ for the decoder $G$ and the discriminator $D$, a feature matching loss \(L_{fm}\), and mel-spectrogram reconstruction loss \(L_{mel}\) from HiFi-GAN:
\begin{equation}
\label{eq2}
\begin{aligned}
&L_G=L_{adv}(G)+\lambda_{fm}L_{fm}+\lambda_{mel}L_{mel},\\
&L_D = L_{adv}(D).
\end{aligned}
\end{equation}
Additionally, to improve computational efficiency and enhance model generalization, we randomly extract segments of latent representations and corresponding audio segments from ground truth raw waveforms as training targets.

\subsection{Inference}
During inference, the model converts speech by replacing the speaker representation. Given a source utterance and a single target utterance, the disentangler extracts content \( C^{\text{src}} \) from the source and speaker identity \( S_{avg}^{\text{tgt}} \) from the target. The decoder then generates the converted waveform $\hat{X}_{\text{conv}}$:
\begin{equation}
    \hat{X}_{\text{conv}} = G(C^{\text{src}} + R_{avg}^{\text{tgt}}).
    \vspace{3pt}
\end{equation}

\section{EXPERIMENTS}
\label{sec:pagestyle}

\subsection{Experimental Setups}
We use the LibriSpeech dataset~\cite{librispeech}, specifically the train-clean-100 subset for training, which consists of 100 hours of speech, and the test-clean subset for evaluation, which includes 5.4 hours of speech from 40 speakers. All audio is downsampled to 16 kHz.
Preprocessing settings include FFT, window, and hop sizes of 1280, 1280, and 320, respectively. KQ is applied using mini batch k-means~\cite{scikit-learn} with a batch size of 2048 and \( K = 256 \) clusters. The speaking variations \( V \) is reduced to $M=8$ dimensions, and the bottlenecked quantized representation \( Q' \) has a dimension of 1016.
The length of the randomly extracted segments, as defined in section~\ref{sec:Training Objective}, is 28.
The encoder is frozen during training, while other modules are optimized with AdamW with $\beta_1 = 0.8$, $\beta_2 = 0.99$, weight decay $\lambda = 0.01$, and an initial learning rate of $2\times10^{-4}$, decaying by a factor of 0.999 each epoch, with a batch size of 32.

\subsection{Baseline Methods}
The proposed method is compared against several state-of-the-art zero-shot VC approaches.
SSR-VC~\cite{ssr-vc} employs HuBERT with a 100-codebook KQ as the encoder with external speaker encoder and incorporates a pitch predictor with a dedicated loss function.  
SEF-VC~\cite{sef-vc} integrates HuBERT with a 2000-codebook KQ and a conformer-based network, incorporating a supervised prosody predictor to enhance prosody modeling.  
kNN-VC~\cite{knn-vc} utilizes WavLM $6^{th}$ layer features, replacing each source feature with the average of its nearest neighbors in a target utterance using k-nearest neighbors (kNN) regression.
Phoneme Hallucinator~\cite{phoneme_hallu} extends kNN-VC by incorporating a VAE-based architecture to generate synthetic target-matching features, improving phonetic coverage in zero-shot scenarios.
DDDM-VC~\cite{dddm-vc} employs WavLM features alongside a diffusion-based network for disentangling speaker identity and content information.

\subsection{Evaluation Metrics}
We conduct both subjective and objective evaluations. In the subjective evaluation, 50 participants rate naturalness and similarity on a 5-point scale using the mean opinion score (MOS) and similarity mean opinion score (SMOS), with 95\% confidence intervals. Experiments are conducted on 20 unseen speakers from the LibriTTS test set.
For the objective evaluation, we used 6 metrics. Intelligibility was assessed via word error rate (WER) and character error rate (CER) calculated with a Whisper-based ASR model\footnote{\href{https://github.com/openai/whisper}{https://github.com/openai/whisper}}~\cite{whisper}. Speaker similarity was measured with equal error rate (EER) and speaker embedding cosine similarity (SECS) using a pre-trained speaker encoder, Resemblyzer\footnote{\href{https://github.com/resemble-ai/Resemblyzer}{https://github.com/resemble-ai/Resemblyzer}}. Prosody preservation was evaluated with the Pearson correlation coefficient (PCC)~\cite{f0-pcc} for both F0 (F0-PCC) and energy (E-PCC), where higher PCC values indicate better prosody preservation. For evaluation, 1000 utterances were randomly selected from the Librispeech test set, and 500 unseen-to-unseen converted utterances are generated.

\section{RESULTS}
\label{sec:pagestyle}
\subsection{Comparison with Baselines}
Table~\ref{table_1} presents the subjective and objective evaluation results comparing the proposed method with baseline models. The proposed model achieves the lowest WER of 5.57\% and CER of 2.07\%, demonstrating superior intelligibility. It also obtains the highest speaker similarity, with a SMOS of 4.27 and a SECS of 0.815, and shows strong prosody preservation with an F0-PCC of 0.719 and an E-PCC of 0.885. Additionally, the MOS results indicate the best naturalness. These results confirm that QR-VC enables high-fidelity zero-shot voice conversion, preserving both content and prosody while enhancing speaker similarity. This validates that the Linear Disentangler ensures distinct separation of speech components, preventing interference or information leakage.

\subsection{Feature Representation Analysis}
\label{sec:4.1}
\begin{figure}[!tb]
\begin{minipage}[b]{1\linewidth}
  \centering
  \centerline{\includegraphics[width=8.9cm]{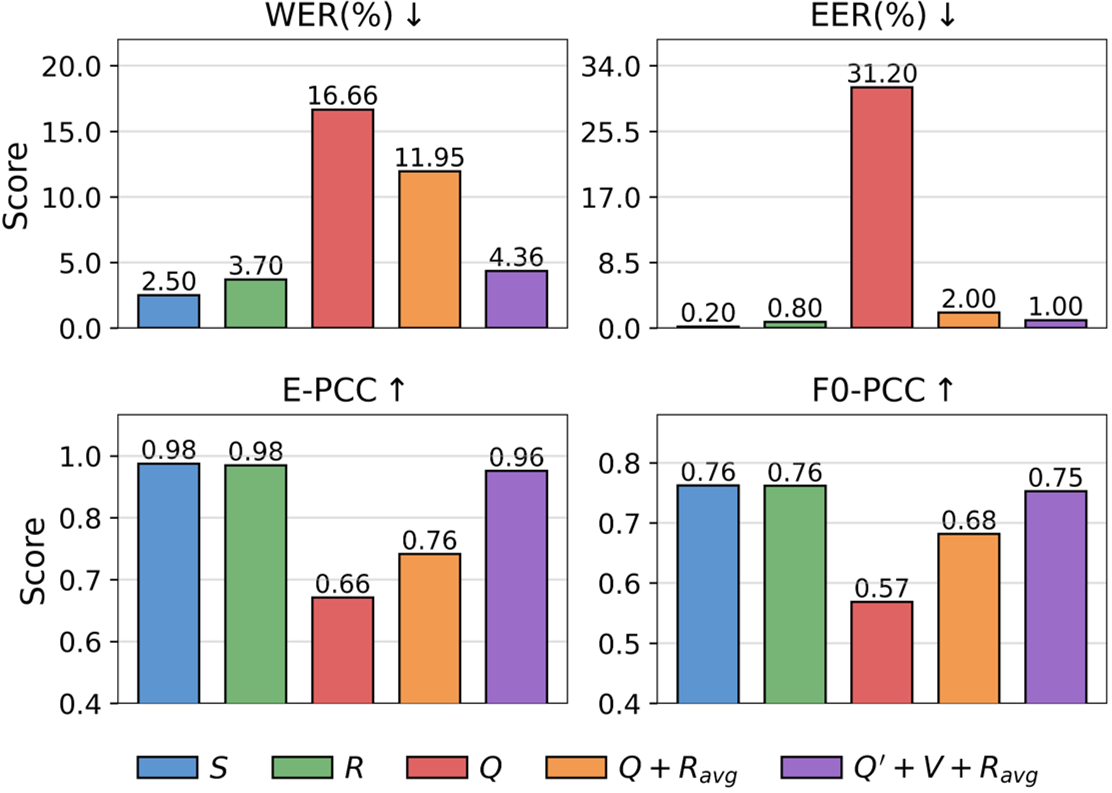}}
\end{minipage}
\caption{Objective evaluation of reconstructed speech from different feature representations.}
\label{feature_analysis}
\end{figure}

We conduct a speech reconstruction task using representations derived from SSL features to evaluate their ability to recover the original waveform.
This experiment assesses how well various speech components are preserved in the reconstructed speech, explicitly analyzing the information retained by each representation.
Figure~\ref{feature_analysis} presents the objective evaluation of reconstructed speech using various feature representations, including SSL features \(S\), quantized representation \(Q\), quantization residual \(R\), bottlenecked quantized representation \(Q'\), speaker identity \(R_{\text{avg}}\), and speaking variations \(V\).

SSL features \(S\) encode rich phonetic, prosodic, and speaker information, resulting in high intelligibility and speaker similarity. In contrast, quantized representation \(Q\) effectively removes speaker identity while preserving essential linguistic content, but also eliminates speaking variations, degrading prosody and content preservation. The quantization residual \(R\) retains fine-grained phonetic and prosodic details, along with speaker identity, suggesting its potential as a valuable feature for speech processing.

The combination of \(Q + R_{avg}\) significantly improves the EER, confirming that \(R_{avg}\) successfully captures speaker identity. Moreover, it enhances reconstructed speech quality, reducing the WER. The combination of \(Q' + V + R_{avg}\) achieves the best trade-off among intelligibility, prosody preservation, and speaker similarity. These results also validate the effectiveness of the Linear Disentangler's ability to separate key speech components while preserving each of them.

\subsection{Effect of Speaking Variations Compensation (SVComp)}

\begin{figure}[!tb]
\begin{minipage}[b]{1\linewidth}
  \centering
  \centerline{\includegraphics[width=9cm]{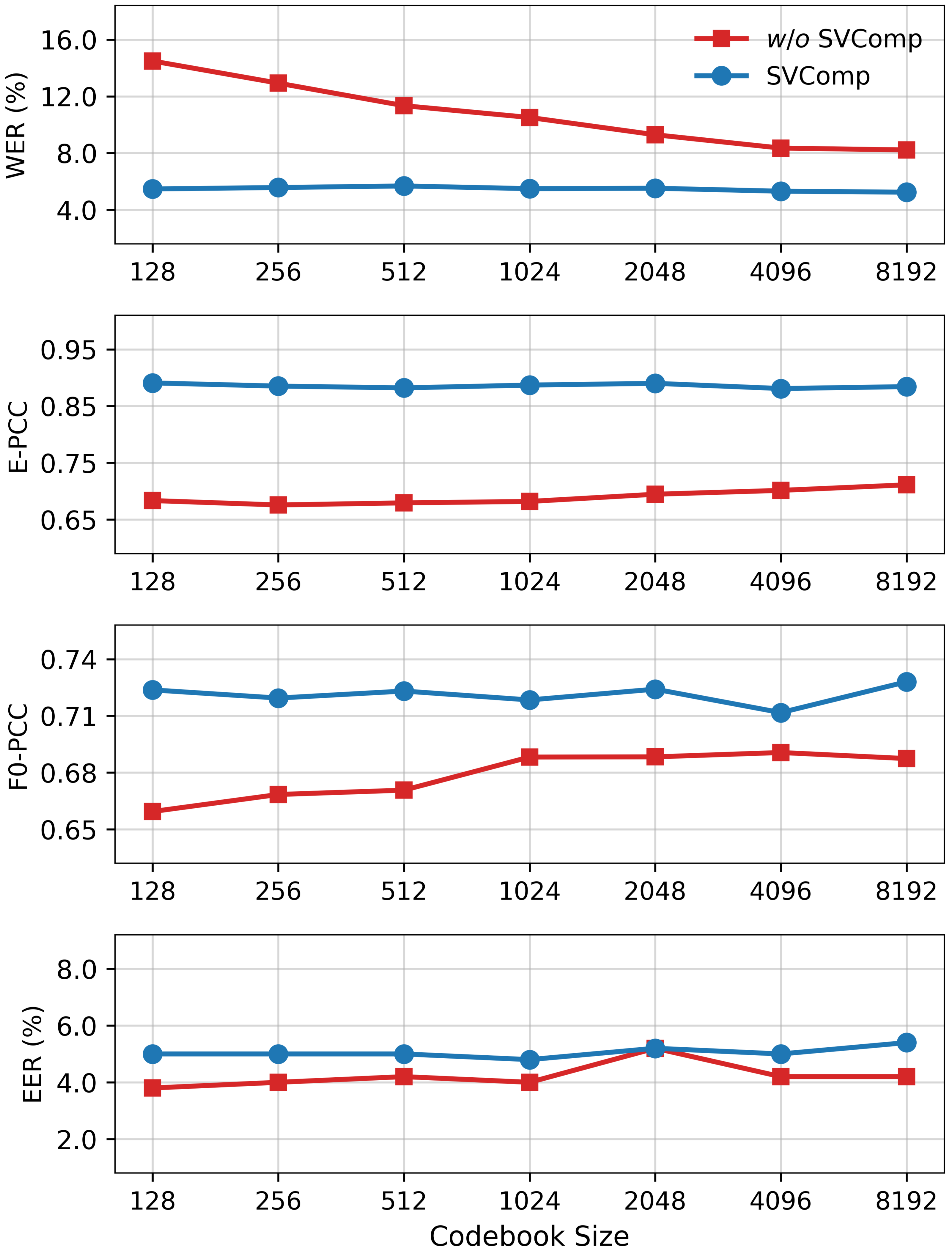}}
\end{minipage}
\caption{Comparison of QR-VC models with and without speaking variations compensation across different codebook sizes.}
\label{exp2}
\end{figure}
Fig.~\ref{exp2} shows the effect of SVComp across different codebook sizes. As the codebook size decreases, the performance gap between models with and without SVComp increases, suggesting that preserving fine-grained phonetic and prosodic details becomes increasingly difficult with lower quantization resolutions. Despite this, the model with SVComp consistently achieves lower WER while maintaining higher prosody consistency, as evidenced by the E-PCC and F0-PCC.

\subsection{Ablation Studies}
An ablation study is conducted to evaluate the contribution of each component in the proposed method, as shown in Table~\ref{table_2}. Removing SVComp ($w/o$ $SVComp$) results in a substantial increase in WER while reducing F0-PCC and E-PCC, confirming its importance in preserving intelligibility and prosody. Eliminating the bottleneck layer in SVComp ($w/o$ $bottleneck$) slightly improves intelligibility and prosody preservation but degrades speaker similarity, demonstrating its role in suppressing speaker information. Replacing the proposed speaker representation \( R_{avg} \) with an external speaker embedding~\cite{spk_enc} ($external$ $spk$ $emb$), in the absence of SVComp due to distribution mismatch, leads to a higher EER, indicating that the proposed method more effectively captures speaker identity.

These results validate the importance of each component in the proposed framework, demonstrating their critical role in preserving intelligibility, prosody, and speaker similarity.

\setlength{\tabcolsep}{2.3pt} 
\begin{table}[!t]
\centering
\normalsize
\caption{Objective evaluation results of ablation systems.}
\begin{tabular}{lcccc} 
\toprule
Model & WER(\%)$\downarrow$ & EER(\%)$\downarrow$ & E-PCC$\uparrow$ & F0-PCC$\uparrow$ \\
\midrule
 Ours & 5.57 & 5.0 & 0.885 & 0.719 \\
\small\textit{ -w/o SVComp} & 12.93 & 4.4 & 0.676 & 0.668 \\
\small\textit{ -w/o bottleneck} & 4.09 & 8.8 & 0.903 & 0.729 \\
\small\textit{ -external spk emb} & 8.55 & 22.6 & 0.836 & 0.664 \\
\bottomrule
\end{tabular}
\label{table_2}
\end{table}

\section{CONCLUSION}
\label{sec:majhead}

This paper proposes QR-VC, a novel zero-shot voice conversion method that leverages quantization residuals in a linear disentanglement framework. By exploiting the temporal properties of speech components, the model disentangles linguistic content, speaker identity, and speaking variations using only K-means quantization and linear projections. This design enables high-fidelity voice conversion relying exclusively on reconstruction losses.
Experiments show that QR-VC outperforms baselines in terms of naturalness, intelligibility, speaker similarity, and prosody preservation. Ablation studies confirm the importance of residual-based speaker representation and the bottleneck structure in disentangling speech components. Evaluations across codebook sizes highlight the effectiveness of the speaking variation compensation module, especially under low-resolution quantization.
Future work includes utilizing the model’s simplicity for real-time and on-device voice conversion.

\section{Acknowledgment}
This research was partly supported by Basic Science Research Program through the National Research Foundation of Korea(NRF) funded by the Ministry of Education(2022R1A6A1A03052954), Institute of Information \& communications Technology Planning \& Evaluation (IITP) grant funded by the Korea government(MSIT) (No.RS-2019-II191906, Artificial Intelligence Graduate School Program(POSTECH)), and the Korea Institute of Energy Technology Evaluation and Planning(KETEP) grant funded by the Ministry of Trade, Industry \& Energy(MOTIE) of the Republic of Korea (No.20214810100010).
\small
\bibliographystyle{IEEEtran}
\bibliography{ref}



\end{document}